\documentclass[aps,prl,onecolumn,superscriptaddress]{revtex4-1}
\usepackage{graphicx}
\usepackage{anysize}
\usepackage{lineno}
\usepackage[amssymb]{SIunits}
\usepackage{lipsum}
\usepackage{amsmath}
\usepackage{amssymb}
\usepackage[toc,page]{appendix}
\usepackage{SIunits}
	\usepackage{soul}
\usepackage{color}
\usepackage{bm}

\newcommand \be {\begin{equation}}
\newcommand \ee {\end{equation}}
\newcommand \bea {\begin{eqnarray}}
\newcommand \eea {\end{eqnarray}}

\begin{document}


\title{Collective orientation of an immobile fish school,	 effect on rheotaxis.}



\author{Renaud Larrieu}
\affiliation{Universit\'e Grenoble Alpes, CNRS, LIPhy, F-38000 Grenoble, France}
\author{Catherine Quilliet}
\affiliation{Universit\'e Grenoble Alpes, CNRS, LIPhy, F-38000 Grenoble, France}
\author{Aur\'elie Dupont}
\affiliation{Universit\'e Grenoble Alpes, CNRS, LIPhy, F-38000 Grenoble, France}
\author{Philippe Peyla}
\email[]{philippe.peyla@univ-grenoble-alpes.fr}
\affiliation{Universit\'e Grenoble Alpes, CNRS, LIPhy, F-38000 Grenoble, France}

\date{\today}

\begin{abstract}
We study  the orientational order of an immobile fish school. Starting from the second Newton's law, we show that the inertial dynamics of orientations is ruled by an Ornstein-Uhlenbeck process. This process describes the dynamics of alignment between neighboring fish in a shoal - a dynamics already used in the literature for mobile fish schools.  Firstly, in a fluid at rest, we calculate the global polarization ({\it i.e.} the mean orientation of the fish) which decreases rapidly as a function of the noise. We show that the faster a fish is able to reorient itself, the more the school can afford to reorder itself for important noise values. Secondly, in the presence of a stream, each fish tends to orient itself and swims against the flow: the so-called rheotaxis. So even in the presence of a flow, it results in an immobile fish school. By adding an individual rheotaxis effect to alignment interaction between fish, we show that in a noisy environment, individual rheotaxis is enhanced by alignment interactions between fish. 
\end{abstract}

\maketitle

\section{Introduction}
The appearance of self-organization within a group of active entities is a fascinating phenomenon \cite{Vicsek2012}. It has been studied for micro-organisms \cite{Levine2000,Clement2019} as well as for active synthetic particles \cite{Paxton2005,lobaskin2013} and at larger scales for animals \cite{Levin2019}. Fascinating self-organisation is also observed within fish shoals \cite{Radakov1973, Pavlov2000, Liao2007, Larsson2012}. Milling and schooling are collective phenomena occuring on scales much larger than an individual fish. This phenomenon has been studied since the begining of the $20^{th}$ century \cite{Parr1927}. The structure and the sensitivity to external factors such as water temperature, light and darkness were analyzed by Breder \cite{Breder1951}. Individual fish in a school were observed to swim for a longer duration when aligned, with lower tail-beat frequencies, smaller energy dissipation and respiratory rates, compared to fish swimming alone \cite{Svendsen2003, Ross1992, Ashrafa2017}. In addition, shoaling and alignment between fish are established as a result of many social and sensory factors like metabolism \cite{Parker1973}, alignment by vision \cite{Partridge1980} or food \cite{Krause1993}. Recently, the study of out-of-equilibrium active systems \cite{Reichhardt2018} allowed scientists to substantially improve their knowledge in modeling this remarkable phenomenon. In the seminal work of Vicsek \cite{Vicsek1995}, an individual animal (bird or fish) adopts instantaneously the average orientation of its neighbors in the group, resulting in a collective motion that can be destroyed by noise. The noise source can be intrinsic to the fish or due to external conditions such as turbulent fluid flow \cite{Liao2007}. Since then, more sophisticated force models have emerged that reproduce quite well the real behavior of schools of fish \cite{Gautrais2009,Gautrais2012,Calovi2014}. That class of social model allows to  study several situations with some flexibility \cite{Calovi2018}. 

If collective motions have been extensively studied \cite{Vicsek2012}, quite poor literature is devoted to immobile groups of fish \cite{Hamilton1971} which stay at the same place relatively to their living environment.  Immobile fish schools can be observed in various situations and especially in reef regions \cite{Keenleyside1979}. Origins of such an immobile state are diverse. It is likely that schools of fish that stop their movements and remain motionless for a period of time may achieve perceptual benefits \cite{Larsson2012}. Simultaneous stopping of a school of fish provides relatively quiet intervals to allow reception of potentially critical environmental signals, fish under predator threat that form non-moving “look around shoals” \cite{Radakov1973} may be an example. However, the most frequent origin of immobile school is a rheotactic effect that allows the fish to orient against a stream \cite{Pavlov2000} and is the object of the present model. This effect was studied in details by Potts \cite{Potts1970}. A school of the {\it snapper Lutjanus monostigma} was observed during several days and self-organized into a polarized and immobile school when submitted to tidal flow. Each fish were heading into the current in order to maintain their position by positive rheotaxis. This is done by swimming gently at an equal and opposite speed to the current. Indeed, by pointing ahead in a direction opposed to the flow can help the school to maintain its immobility in a region where food is present.   A fish can individually find the direction of flow through sentitive captors \cite{Montgomery1997, Kulpa2015} and can also try to align with its congeners. 

In the following, we will first present the model of fish orientation with respect to neighbours and flow. We then show that alignment interactions within a shoal can increase rheotaxis efficency of a single fish.

\section{Model}
\begin{figure}[htbp]
	\centering
	\includegraphics[width=8cm]{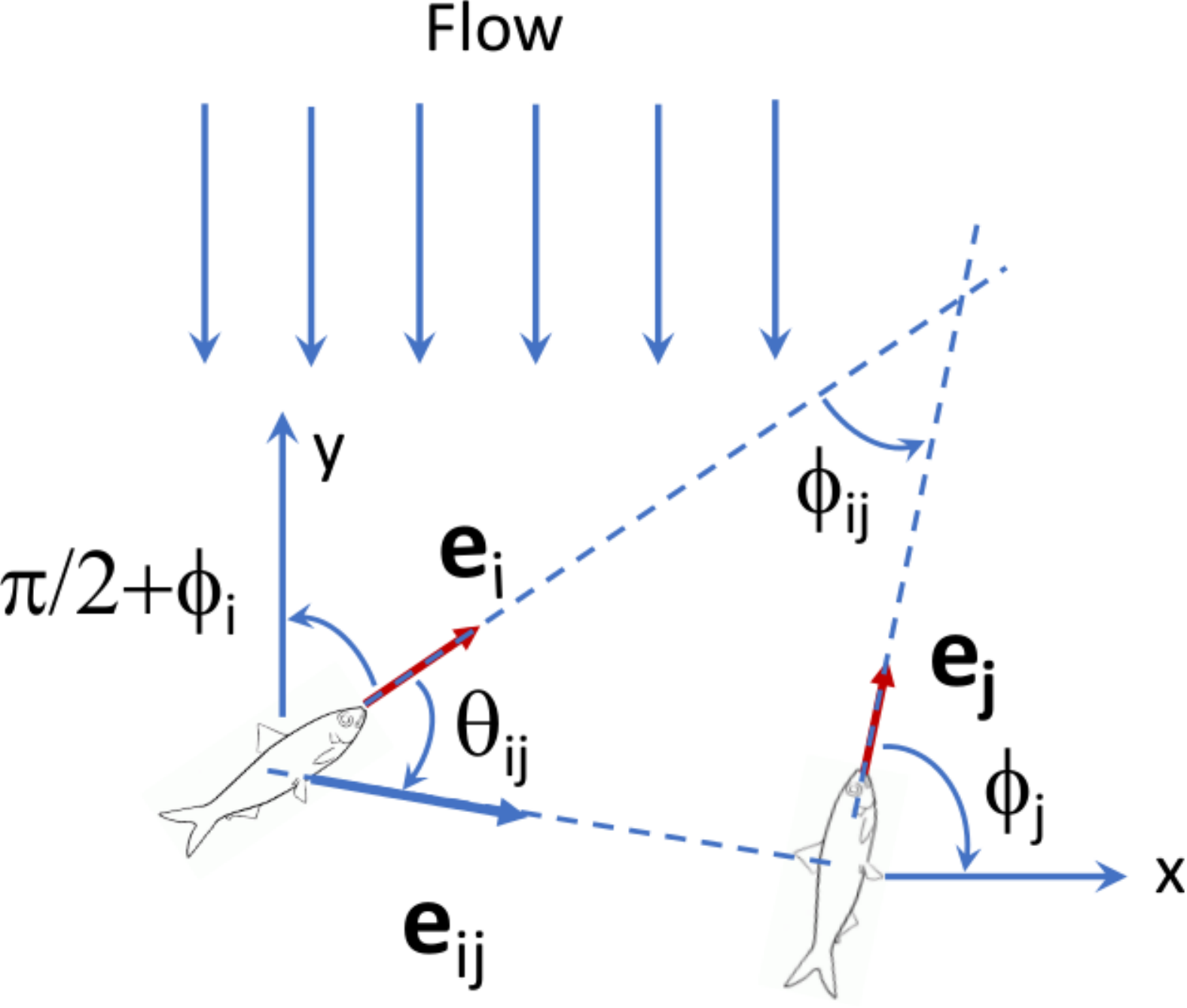}  	   
	\caption{Fish in a flow. The flow is oriented toward the $y<0$-direction. 
	Angles: $\phi_i$ is the angle between orientation $\mathbf{e}_i$ of the $i^{th}$ fish and the horizontal $x$-axis; $\theta_{ij}$ is the angle between fish $i$ orientation $\mathbf{e}_i$ and the fish $i-j$ orientation vector $\mathbf{e}_{ij}$; finally $\phi_{ij}$ is the angle between orientaions of fish $i$ and $j$.}  \label{scheme}
\end{figure}
 Let's consider a motionless fish (fig.\ref{scheme}) located at a fixed position and with a time-varying orientation, living in a school and interacting with its neighbors while attempting to orient itself in a direction opposite to a uniform flow (rheotaxis). Here, for simplicity we use a 2D fish shoal \cite{Gautrais2012,Calovi2014} with a circular shape of radius $R$. 
 Each of the $N$ fish is a discoid of radius $a$. When we vary $N$, we maintain the density $\rho=N (a/R)^2$ constant, typically, $\rho \approx 0.5$. This corresponds to a quite dense school which can be oftenly encontered \cite{Simmonds2008} and which allows us to have small fluctuations of the local density of fish.
 In the spirit of social force model originally developped by Helbing \cite{Helbing1995,Helbing2000}, we consider here social torques. These "torques" are a measure for the internal motivations of each individual to perform certain movements (rotation) depending on its environment. Each fish needs to orient itself in the same direction as its neighbors within a chosen radius of $5a$ ({\it i.e.} 2.5 times its own size) and against the flow. The typical size $5a$ has been chosen to capture neighbors that are in the close neighboring of a given fish regarding the chosen density $\rho$. Starting with the second Newton's law for rotating bodies we can write:
 \begin{equation}
 \mathcal{I}\dot{\omega}_i=-\zeta \omega_i+\sum_{j \in V_i}T^I_{ij}(\phi_{ij})+T^R_i(\phi_i)+\eta(\mathbf{r}_i,t),
 \label{Newton}
 \end{equation}
 where $\mathcal{I}$ is the moment of inertia of any fish (supposed identical), $\omega_i$ is the angular velocity of fish $i$ located in $\mathbf{r}_i$ and oriented with an angle $\phi_i$ with the $x$-axis at time $t$. The angular acceleration is $\dot{\omega}_i$. Fluid friction is $\zeta$. The $i^{th}$ fish interacts with its close congeners indexed by $j$ in the $V_i$ ensemble of its $N_i$ neighbors ({\it i.e.} within a circle of radius $5a$ around fish $i$). This interaction is represented by a social torque $T^I_{ij}$ which depends on the relative orientation between fish $i$ and $j$: $\phi_{ij}=\phi_i-\phi_j$. The torque associated with the rheotaxis is $T^R_i$ which depends on the orientation of the fish $i$ with the $y$-axis chosen as the direction of the incident and uniform flow. Finally, the dynamics is perturbed by a noise term $\eta(\mathbf{r},t)$ with $<\eta(\mathbf{r},t)>=0$ and $<\eta(\mathbf{r},t)\eta(\mathbf{r'},t')> \sim \delta(\mathbf{r}-\mathbf{r'})\delta(t-t')$. Note that in our model, we do not consider an interaction term that depends explicitly on fish interdistance.
 This equation can be rewritten as:
 \begin{equation}
 \dot{\omega}_i=-\frac{1}{\tau} \omega_i+\frac{1}{\tau}\omega_i^*(\phi_{ij})+\eta(\mathbf{r}_i,t)/\mathcal{I},
 \label{OU1}
 \end{equation}
 which expresses that each fish adjusts its angular velocity $\omega(t)$ towards a time-dependent target value $\omega_i^*=\omega_i^{I*}+\omega_i^{R*}$, depending both on the fish-fish interaction ($\omega_i^{I*}$) and on the rheotaxis ($\omega_i^{R*}$) within an external flow. Expressions of $\omega_i^{I*}$ and $\omega_i^{R*}$ are given below. We use the time $\tau$ associated with dissipation  as the characteristic time and we rescale by $\tau$ the other times associated with alignments and rheotaxis. Note that if we consider a fish with a size around a few $cm$, the time $\tau$ associated with dissipation during a solid rotation is about a few seconds which is much bigger than a typical time of reaction for alignment  closer to a few tenths of a second. However, the  time associated with dissipation can be much shorter if we consider that usually a fish is flexible and a change of orientation is driven by a deformation of its body which can reduces very much its inertia \cite{Porter2011}.  Following the spirit of \cite{Gautrais2012,Calovi2014} in order to describe the interaction between close fish, we write :
 \begin{equation}
 \omega_i^{I*}=\frac{I}{\tau}\,\frac{1}{N_i}\sum_{j \in V_i} \sin(\frac{\phi_{ij}}{2})\,\frac{1+\cos(\theta_{ij})}{2},
 \label{omegastarinterac}
 \end{equation}
 where $I$ is the dimensionless amplitude of alignment interaction. The term $\sin(\phi_{ij}/2)$ accounts for alignment between fish $i$ and $j$. If fish $i$ and $j$ are aligned in the same direction, it reads $\sin(\phi_{ij}/2)=0$ and then $\omega_i^*=0$ since fish $i$ has no reason to rotate. But if fish $i$ and $j$ are anti-parallel, {\it i.e.} $\phi_{ij}=\pm \pi$, then $\sin(\phi_{ij}/2)=\pm 1$, since fish $i$ must rotate. The term $\left ( 1+\cos \theta_{ij} \right )/2$ is designed to ensure a frontal preference and some kind of rear blind angle \cite{Calovi2014}.
 In order to model the alignment against the flow (rheotaxis), we have:
 \begin{equation}
 \omega_i^{R*}=\frac{F}{\tau}\sin \left(\frac{\pi/2+\phi_i}{2} \right).
 \label{rheotaxis}
 \end{equation}
 The term $F$ is dimensionless and represents the amplitude of the rheotaxis.  If $\phi_i=-\pi/2$, the fish does not rotate since it is already aligned against the flow ({\it i.e.} pointing in the $y>0$ direction) it reads $\omega_i^{R*}=0$. But if the fish is aligned along the flow, ($\phi_i=+\pi/2$) the fish must turn back in order to point against the flow with the target angular velocity $\omega_i^{R*}=F/\tau$.

 Using an Euler-Maruyama integration \cite{Higham2001}, eq.(\ref{OU1}) reads:
   \begin{equation}
  d\omega_i(t)=-\frac{dt}{\tau} \left[ \omega_i(t)-\omega^*_i(t) \right]+\sigma \mathcal{N}\sqrt{dt},
  \label{OU2}
  \end{equation}
  also known as an Ornstein-Uhlenbeck process \cite{Gautrais2009, Gautrais2012}. The noise amplitude is $\sigma$ and $\mathcal{N}$ is a random gaussian variable of mean $0$ and variance $1$. For large values of time, this equation becomes stationnary (see appendix). In the absence of rheotaxy ($F=0$),
 we can rescale time and angular velocities by $\tau/I$ and $I/\tau$ respectively.  At stationnarity, it is easy to show that there is only one dimensionless  number  $\tilde{\sigma}=\sigma \tau^{3/2}/I^{1/2}$ that compares the amplitude of noise and the amplitude of alignment interaction (see appendix).  
 
 In the presence of rheotaxis ($F \neq 0$), three terms should be compared: the alignment interaction (amplitude $I$), the rheotaxis (amplitude $F$) and the noise term (amplitude $\sigma$). Since we would like to vary $F$ at constant $I$, we choose to rescale time and angular velocities by $\tau/F$ and $F/\tau$ respectively. Then, we get two dimensionless numbers $I/F$ and $F^{1/2}/(\sigma \tau^{3/2})$. The last term being equal to  $(F/I)^{1/2}/\tilde{\sigma}$.
 
 In the following, we will integrate numerically eq.(\ref{OU2}) in the absence or in presence of rheotaxis. For each situation, we will plot the global polarization (defined below) as a function of the above dimensionless numbers.

 \begin{figure}
	\includegraphics[width=8cm]{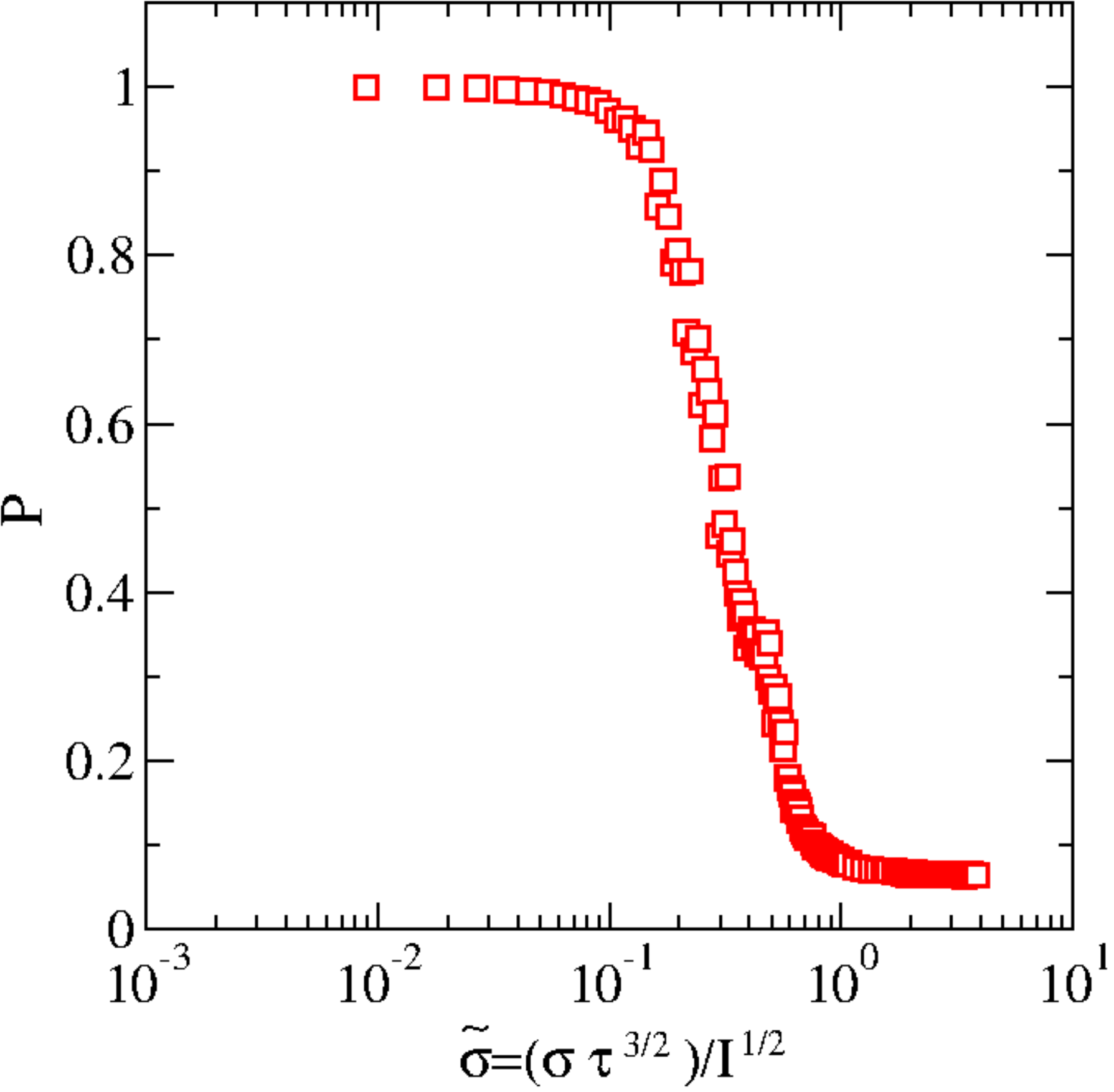}	   
	\caption{Decrease of the global polarization as a function of the dimensionless noise $\tilde{\sigma}$ without rheotaxis ($F=0$). Each point is averaged over $20$ runs. The total time of each simulation is $t_{max}/(\tau/I) = 10^4$ with a time step $dt/(\tau/I) = 10^{-2}$.}  \label{polarization}
	
\end{figure}

 \section{Results.}

\begin{figure}
	\centering
	\hspace*{-0.50cm}
	\includegraphics[width=10cm]{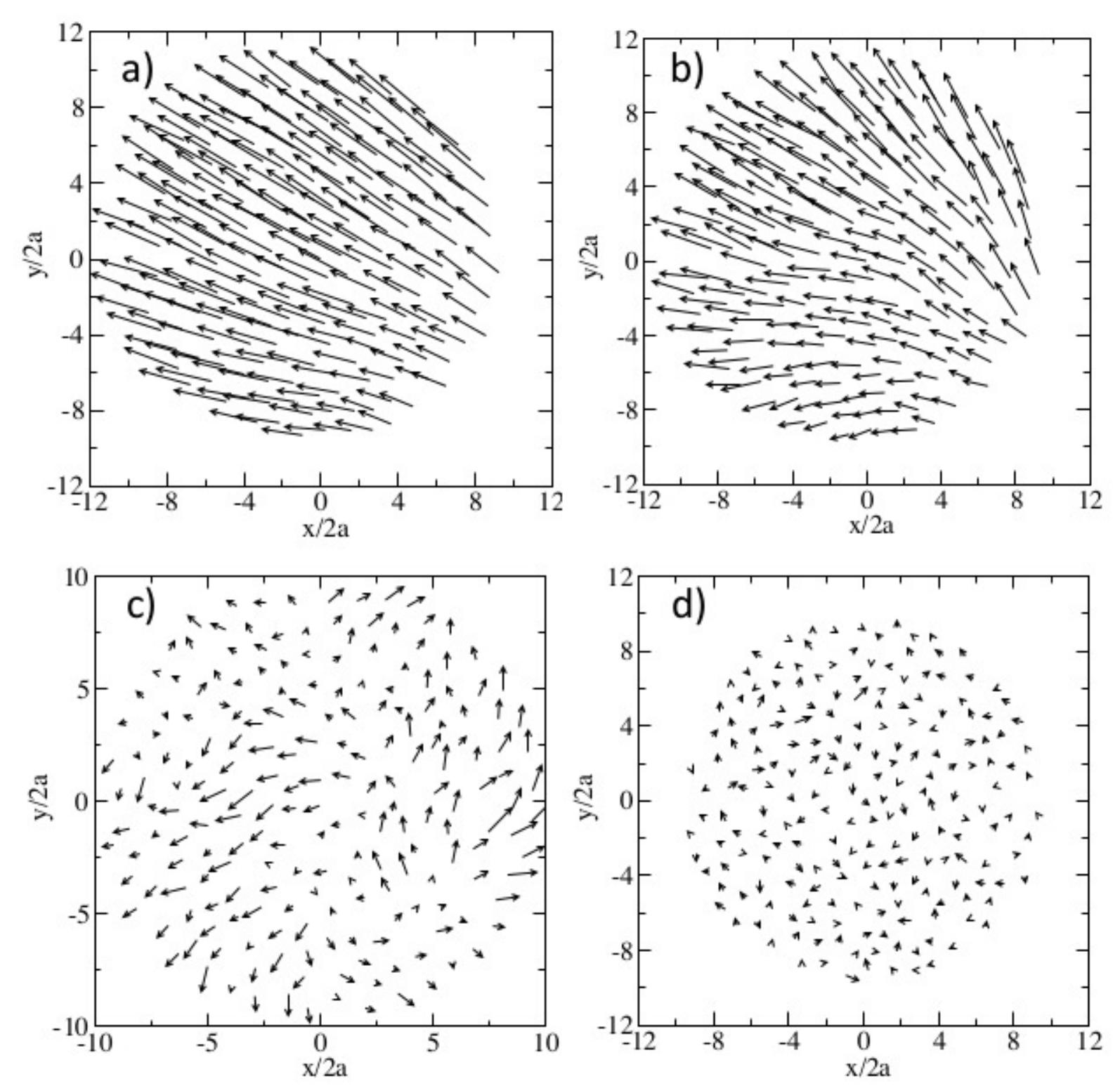}  	   
	\caption{Time averaged orientations of each fish for different values of the noise without rheotaxis ($F=0$). A short or large arrow indicates small and large polarization of a given fish respectively. a) For $\tilde{\sigma}=0.18$, fish are well polarized. b) For $\tilde{\sigma}=0.27$, a small zone of weak polarization appears. c) For a larger value $\tilde{\sigma}=0.54$, only a few islands of polarized fish remain. d) Then, for $\tilde{\sigma}=1.0$, no polarization remains (time averaged value $\bar{\mathbf{e}}_i \approx 0$).}   \label{orientation}
\end{figure}

  We first analyse the alignment within an immobile school of $N=200$ fish as a function of the noise $\tilde{\sigma}$ and without rheotaxis ($F=0$). Note that we tried different numbers of fish: $N=400$, $N=200$, $N=100$ and $N=50$ (see appendix). Since results are quite close for $N=400$ and $N=200$, we choose to work with $N=200$ fish for this work. 
 
 We compute the mean value of the global polarization \cite{Vicsek1995,Gautrais2012} defined as 
 $P=\frac{1}{N}\left \vert \sum_{i=1}^{N}  \mathbf{e}_i \right \vert$:
 \begin{equation}
 P=\frac{1}{N} \sqrt{ \left( \sum_{i=1}^{N}\cos \phi_i(t) \right)^2 + \left( \sum_{i=1}^{N}\sin \phi_i(t) \right)^2 }.
 \label{polardef1}
 \end{equation}
 The global polarization of the school is $P=1$ when all fish point in the same direction while $P=0$ means that the fish point in different directions.
 
 The school initial orientation is polarized in a given random direction.  After a transient time of order $\tau/I$, fish can loose partially their mutual alignment because of the noise. Thus, we measure $P$ for large values of time ($t \gg \tau/I$) as a function of $\tilde{\sigma}$ (see fig. \ref{polarization}).
 We found that $P$ drops abrupltly for $\tilde{\sigma} \sim 0.27$. Here, the rescaled noise being $\tilde{\sigma}=\sigma \tau^{3/2}/\sqrt{I}$, it means that, fast reacting fish (small values of $\tau/I$) are more able to line up in a stronger noisy environment (large value of $\sigma$) than fish with a larger $\tau/I$ value that cannot sustain the same amplitude of noise to form a polarized school. Note also that when $P=0.6 \pm 1$, we obtain a maximum of polarization fluctuations whatever $N$ (see appendix - fig. \ref{fluctuations}).

\begin{figure}
	\centering
	\hspace*{-0.50cm}
	\includegraphics[width=10 cm]{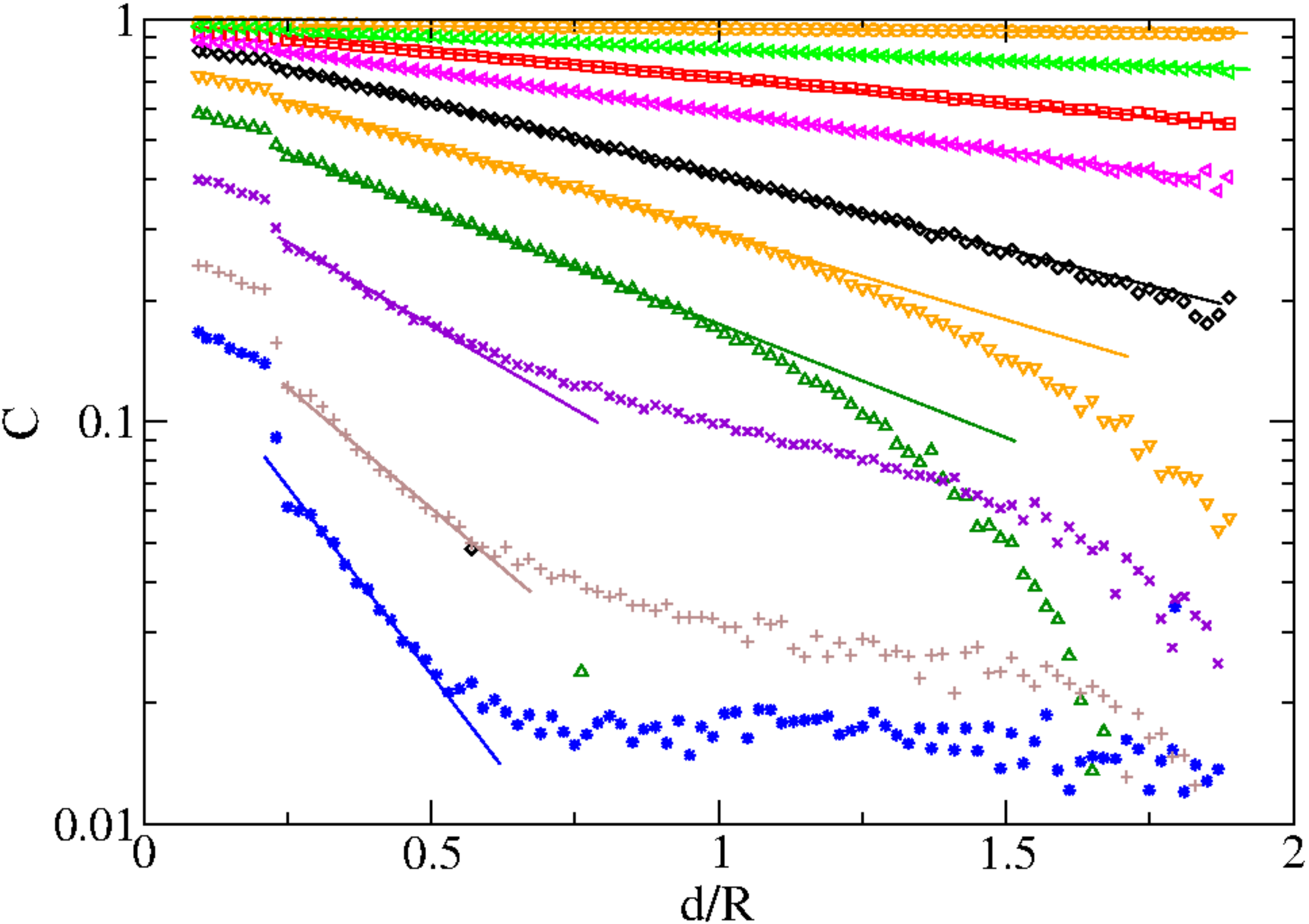}  	   
	\caption{Correlation function $C$ as a function of the fish-to-fish distance $d$ rescaled by $R$. The first part of the curves is associated with the correlation between close neighbor fish ($d<5a$). Then, for $d>5a$, an exponential decay is oberved. Orange circles: $\tilde{\sigma}=0.09$; green left triangles: $\tilde{\sigma}=0.13$ red squares: $\tilde{\sigma}=0.18$; magenta down triangles: $\tilde{\sigma}=0.22$; black diamonds: $\tilde{\sigma}=0.27$; orange down triangles: $\tilde{\sigma}=0.36$; green up triangles: $\tilde{\sigma}=0.45$; purple crosses: $\tilde{\sigma}=0.54$; brown plus: $\tilde{\sigma}=0.63$; blue stars: $\tilde{\sigma}=0.72$. Each solid line is a fit with an exponential decay such as $~exp(-d/\lambda)$, where $\lambda$ is a correlation length (fitting parameter).}  \label{correl}
\end{figure}

\begin{figure}
	\centering
	\hspace*{-2.50cm}
	\includegraphics[width=10 cm]{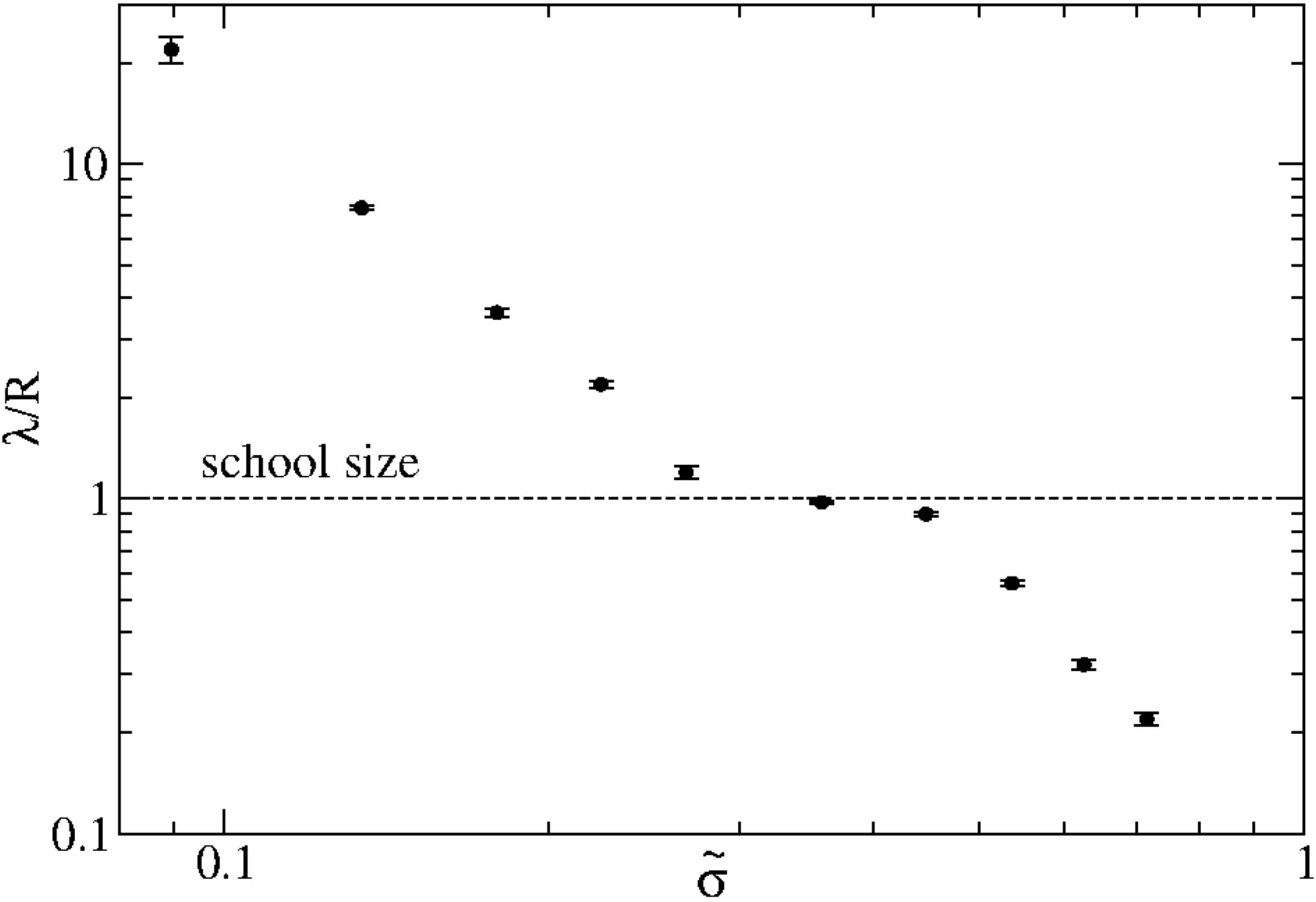}  	   
	\caption{Correlation length $\lambda$ as a function of noise $\tilde{\sigma}$. $\lambda$ decreases as a function of $\tilde \sigma$ and reaches a plateau when becoming comparable to the school size $R$ around $\tilde{\sigma} \approx 0.27$. Above this noise it continues to decrease below $R$.}  \label{correl_length}
\end{figure}

To visualize the loss of orientation when noise increases, let us first consider the map of individual polarizations around $\tilde{\sigma}=0.27$. In fig.\ref{orientation}, we plot the time averaged value of each direction $\bar{\mathbf{e}}_i(t)=1/t_{max}\int_0^{tmax}\mathbf{e}_i dt$ for different values of $\tilde{\sigma}$. We use $t_{max}/(\tau/I)=10^4$ and $dt/(\tau/I)=10^{-2}$.  Starting with $\tilde{\sigma}=0.18$, the group of fish is well polarized (fig. \ref{orientation}.a). Then, we increase the rescaled noise value. We see that for $\tilde{\sigma}=0.27$, (fig. \ref{orientation}.b) a weak polarized region appears. Around $\tilde{\sigma}=0.54$ (fig. \ref{orientation}.c) several weak polarized zones have invaded the school, leading to islands of polarized fish separated by unpolarized zones. For larger values $\tilde{\sigma}=1.0$ (fig. \ref{orientation}.d), an entire unpolarized fish school (a so-called shoal)  remains.  We also calculate the correlation function $C=\overline{\mathbf{e}_i . \mathbf{e}_j}=1/t_{max}\int_0^{tmax}\mathbf{e}_i . \mathbf{e}_j dt$ for different distances $d$ between fish $i$ and $j$. As shown in figure \ref{correl}, $C$ decreases exponentially as a function of the fish-to-fish distance with a typical correlation length $\lambda$. This correlation length decreases by increasing the noise (fig.\ref{correl_length}). We observe a small plateau for $\lambda$ close to  the school size $R$ probably due to a boundary effect.

\begin{figure}
	\centering
	\hspace*{-0.50cm}
	\includegraphics[width=10 cm]{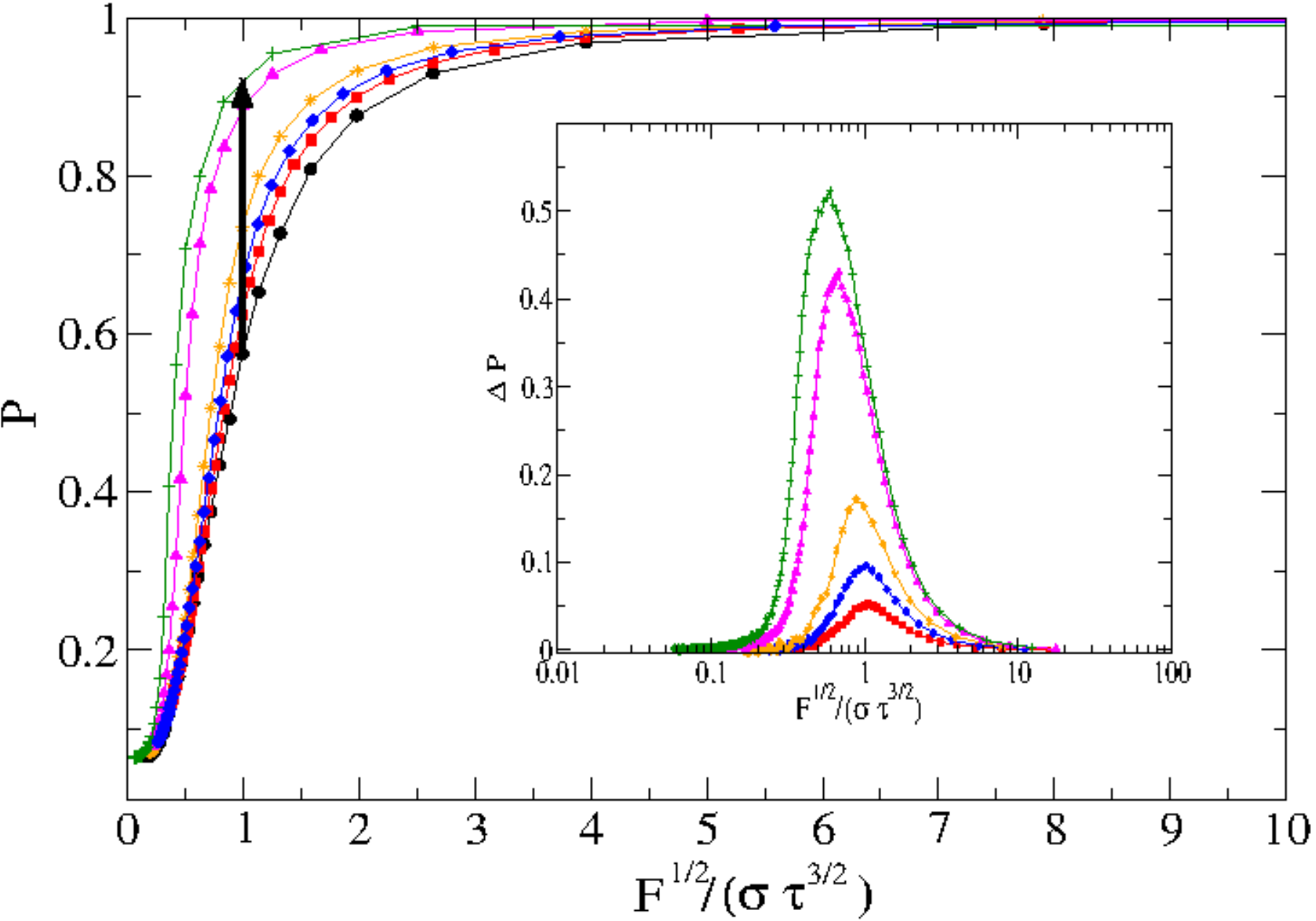}  	   
	\caption{Polarization $P$ of the school against the flow as a function of $F^{1/2}/(\sigma \tau^{3/2})$ for different values of the parameter $I/F$.  $I/F=20.0$ (green crosses), $I/F=10.0$ (magenta triangles), $I/F=2.0$ (orange stars), $I/F=1.0$ (blue diamond), $I/F=0.5$ (red squares), $I/F=0$ (black circles). A significant increase of $P$ is observed when $I/F$ increases. The arrow indicates the polarization difference between $I/F=20.0$ and $I/F=0$ at  $F^{1/2}/(\sigma \tau^{3/2})=1$. Inset:  Polarization difference of the school $\Delta P=P(I/F)-P(I/F=0)$. The colors and symbols represent the same $I/F$ values as on the main figure. The presence of alignment interaction is very efficient for rheotaxis when  $I/F$ is large and $F^{1/2}/(\sigma \tau^{3/2}) \approx 1$.}  \label{diff_F}
\end{figure}

 To study the effect of the alignment interaction between fish on the rheotaxis of the whole group, we now consider non-zero values of $F$. In the absence of alignment interactions between individuals ($I=0$), each fish tends to orient itself against the flow (pointing toward $y>0$ direction). The presence of noise perturbs the rheotactic behavior of each fish and the polarization drops by increasing the noise $\sigma$ or $\tau$ and decreasing the rheotaxis $F$ (fig.\ref{diff_F}). We assume that even if a given fish is not perfectly oriented against the flow it still maintains its position within the shoal in order to stay with its congeners.

  Now, by switching on the alignment interaction between fish ({\it i.e.} for non-zero and positive values of $I$), we observe a clear increase of the global polarization (see figure \ref{diff_F} (black arrow)). Note that the $x$-axis ($\sqrt{F}/(\sigma \tau^{3/2})$) is inversely proportional to $\sigma$.
    In the inset of fig.\ref{diff_F}, we plot the polarization difference $\Delta P=P(I/F)-P(I/F=0)$ between the global polarization $P$ of the school in presence of interactions and in the absence of social interaction ($I=0$).  We see that for large values of $I/F$, a strong increase of the polarization against the flow is obtained and reaches a maximum around $F^{1/2}/(\sigma \tau^{3/2}) \approx 1$.  For small rheotaxis or strong noise the global polarization drops to zero since the fish are pointing in all directions. On the contrary, for large rheotaxis or small noise  all the fish are pointing in the direction opposed to the flow and $P$ saturates. In both cases, the role of fish-fish interaction is inefficient. But between these two extreme cases ($F^{1/2}/(\sigma \tau^{3/2}) \approx 1$), we observe a maximum of $\Delta P$ corresponding to a significant gain of rheotaxy by the interplay of the fish to fish interactions. 
It can be concluded that  \textbf {a strong fish to fish interaction improves the collective rheotaxis even if the individual rheotaxis is weak}.

\section{Conclusion}

In this paper, we have studied the collective orientation of an immobile group of fish with
two ingredients: a social torque to align fish with their close neighbors and an environmental
torque to align fish with an external flow. We have modeled the inertial dynamics of groups of
fish in the presence of noise. In the absence of an external flow, we show that for large values of
a dimensionless noise $\tilde \sigma \gg 1$, the group cannot globally polarize. In the presence of a flow, we show that
strong social interactions help the group to detect and align even with weak individual rheotaxis. 
This model can be extended
to the case of moving fish which for certain species are able to detect low gradients of velocities
(in a Poiseuille flow) \cite{Oteiza2017}. In this case, it would be interesting to study whether collective social 
interactions can improve the efficiency of this specific rheotactic behavior. Despite the
simplicity of the model which does not account for hydrodynamic drag, we believe that several dynamics
of group organization (or collective behaviors) in a complex environment can be captured.

\section*{Acknowledgements}
We thank the Mission Interdisciplinaire du CNRS for a financial support. We thank our colleagues P. Moreau and P. Ballet for their technical help.

\section*{Appendix.}

\subsection{Different numbers of fish.}

\begin{figure}[h]
	\centering
	\hspace*{-0.50cm}
	\includegraphics[width=10 cm]{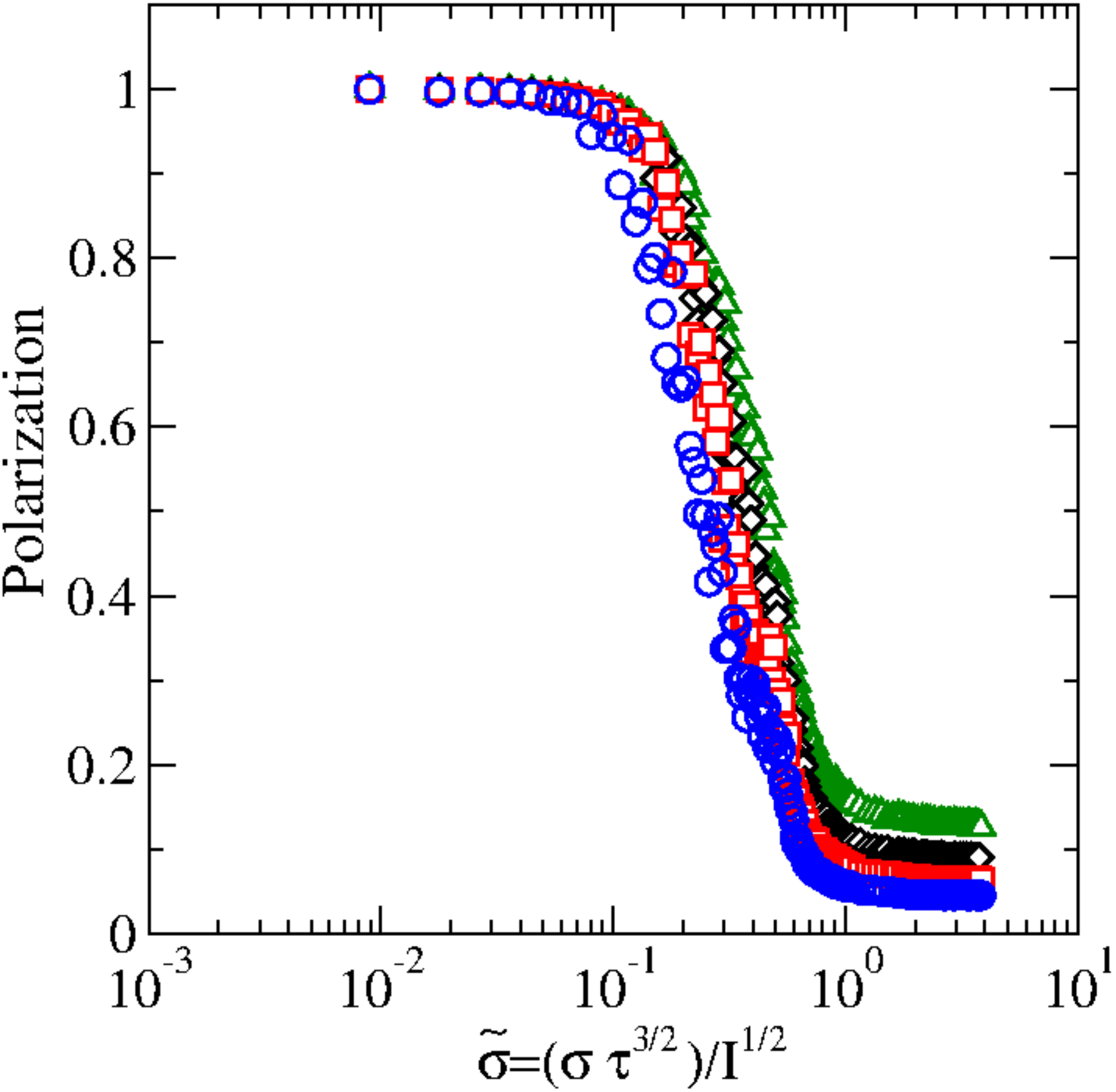}  	   
	\caption{Polarization of the school as a function of noise $\tilde{\sigma}$ for different numbers of fish $N$. Green triangles $N=50$; black diamonds $N=100$; red squares $N=200$; blue circles $N=400$.}  \label{polarNfish}
\end{figure}

\begin{figure}
	\centering
	\hspace*{-0.50cm}
	\includegraphics[width=10 cm]{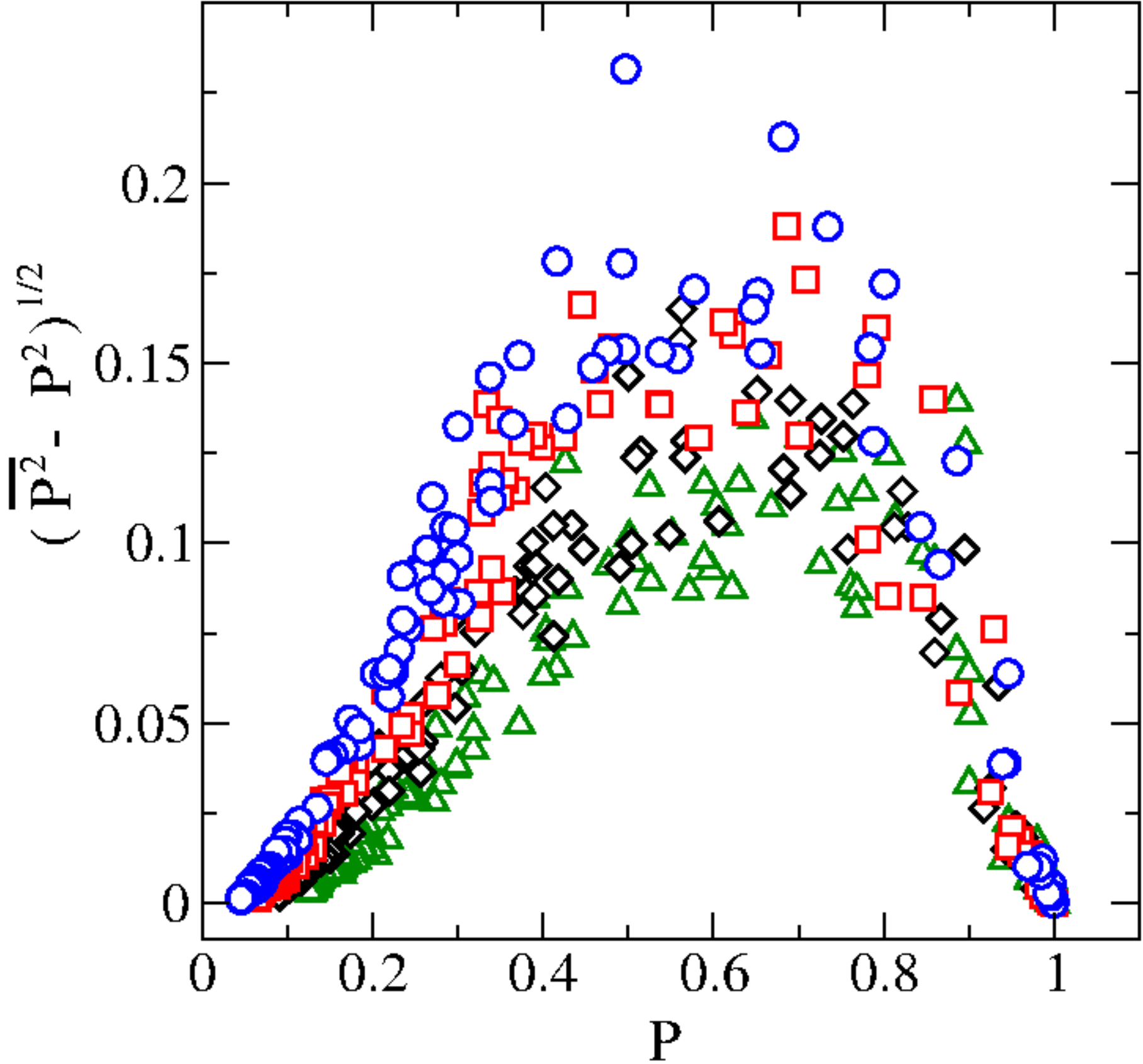}  	   
	\caption{Fluctuation of the school polarization as a function of polarization for different numbers of fish $N$. Green triangles $N=50$; black diamonds $N=100$; red squares $N=200$; blue circles $N=400$. Whatever $N$, the fluctuation maximum is around $P=0.6 \pm 0.1$.}  \label{fluctuations}
\end{figure}

We vary the number of fish, from $N=50$ to $N=400$ (see figure \ref{polarNfish}). Depending on $N$, the drop of polarization  occurs in the range of noise $0.25<\tilde{\sigma} < 0.5$, however the amplitude of fluctuation always occurs around $P=0.6 \pm 0.1$  (Fig.\ref{fluctuations}). By changing $N$, we maintain the density $\rho=N/(\pi R^2)$ constant and equal to $\rho \approx 0.5$. 

\subsection{Dimensionless numbers}
In the absence of rheotaxis ($F=0$) the rescaling of eq.(\ref{OU2}) at stationnarity (see section C) leads to
\begin{equation}
0=-\tilde{dt} \left[ \tilde{\omega}_i(t)-\tilde{\omega}_i^*(t) \right]+\frac{\sigma \tau^{3/2}}{I^{1/2}}\mathcal{N}\sqrt{\tilde{dt}},
\label{OUwithoutFlow}
\end{equation}
where $\tilde{t}=tI/\tau$ and $\tilde{\omega}=\omega \tau/I$. The dimensionless target is:
\begin{equation}
\tilde{\omega}^*_i=\frac{1}{N_i}\sum_{j \in V} \sin(\frac{\phi_{ij}}{2})\,\frac{1+\cos(\theta_{ij})}{2},
\label{targeti}
\end{equation}

Now, in presence of rheotaxis ($F\neq0$) the rescaling of eq.(\ref{OU2}) at staionnarity leads to
 \begin{equation}
0=-\tilde{dt} \left[ \tilde{\omega}_i(t)-\tilde{\omega}_i^*(t) \right]+\frac{\sigma \tau^{3/2}}{F^{1/2}}\mathcal{N}\sqrt{\tilde{dt}},
\label{OUwithFlow}
\end{equation}
where $\tilde{t}=tF/\tau$ and $\tilde{\omega}=\omega \tau/F$. The target is:
 \begin{equation}
\tilde{\omega}^*_i=\frac{I}{F}\frac{1}{N_i}\sum_{j \in V} \sin(\frac{\phi_{ij}}{2})\,\frac{1+\cos(\theta_{ij})}{2}-\sin(\frac{\pi/2+\phi_i}{2}),
\label{targetir}
\end{equation}

In this case, two dimensionless numbers are to be considered $I/F$ and $\sqrt{F}/(\sigma \tau^{3/2}$).

\subsection{Stationnarity}

\begin{figure}
	\centering
	\hspace*{-0.50cm}
	\includegraphics[width=8 cm]{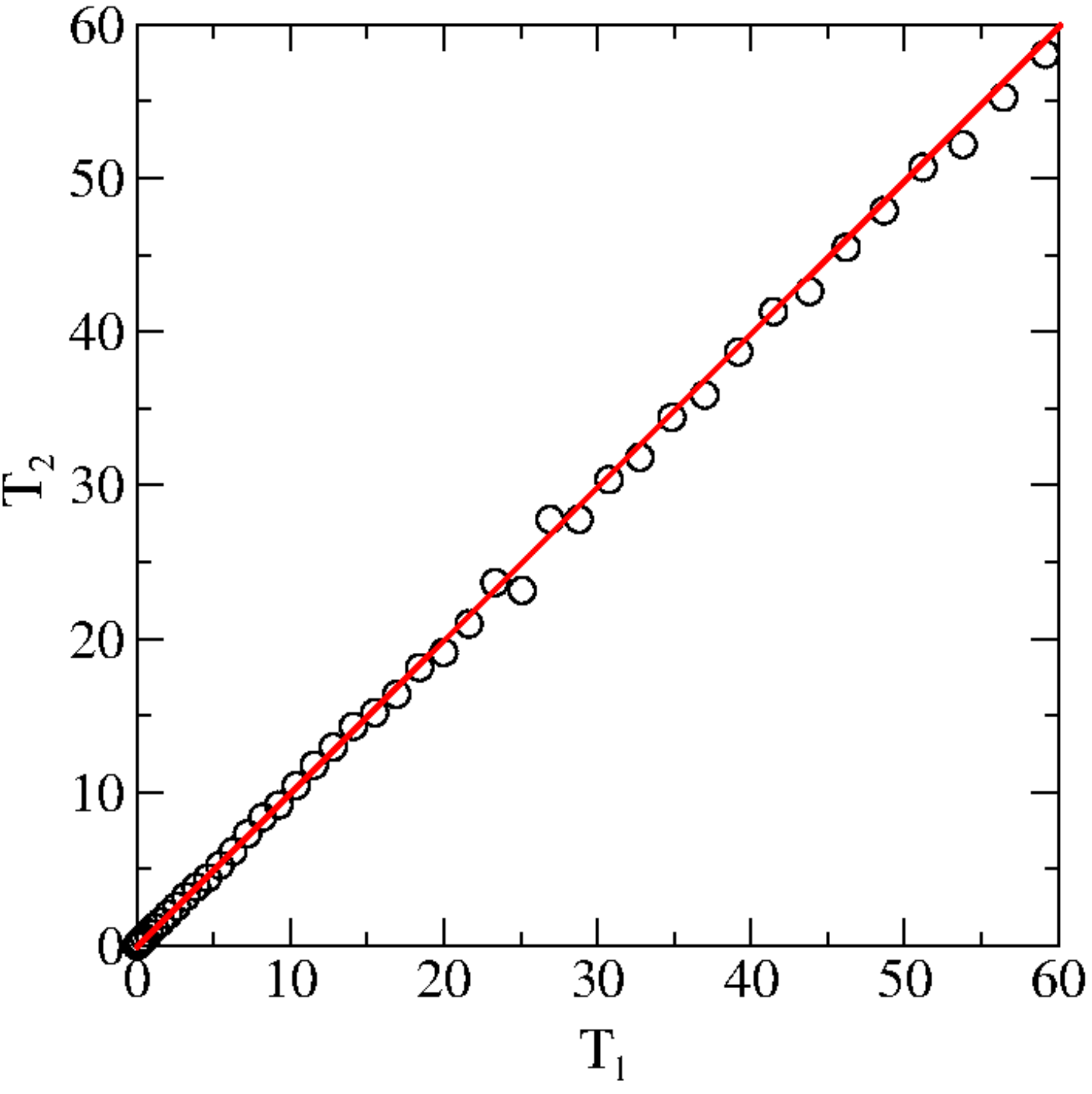}  	   
	\caption{Term $T_1$ ($lhs$ of eq.(\ref{OUstationnarwithFlow2})) as a function of $T_2$ ($rhs$ of eq.(\ref{OUstationnarwithFlow2})), Data are circles and the red solid line represents the first bisector, showing that $T_1=T_2$.}  \label{T1egalT2}
\end{figure}

For large values of time ($\tilde{t} \gg 1$), we can assume stationnarity.  In the presence of rheotaxis, if this hypothesis is true, we should have:
 \begin{equation}
\left[ \tilde{\omega}_i-\tilde{\omega}_i^* \right] \tilde{dt} =\frac{\sigma \tau^{3/2}}{F^{1/2}}\mathcal{N}\sqrt{\tilde{dt}}.
\label{OUstationnarwithFlow}
\end{equation}
But this equality is not easy to prove numerically because of the presence of noise. So let us average each member of eq. (\ref{OUstationnarwithFlow}) on time and fish.  
By integrating on time and using Ito isometry we obtain the following equality:
\begin{equation}
\left\langle \left \{ \frac{1}{\tilde{t}_{max}} \int_{0}^{\tilde{t}_{max}}\left[ \tilde{\omega}-\tilde{\omega}^* \right ]\tilde{dt} \right \}^2 \right\rangle = \left(\frac{\sigma \tau^{3/2}}{F^{1/2}} \right)^2
\label{OUstationnarwithFlow2}
\end{equation}
with $\tilde{t}_{max}=t_{max}/(\tau/F)=10^4$ and $<.>$ represents the averaging over the $N$ fish.
Let us call $T_1=\left\langle \left \{ \frac{1}{\tilde{t}_{max}} \int_{0}^{\tilde{t}_{max}}\left[ \tilde{\omega}-\tilde{\omega}^* \right ]\tilde{dt} \right \}^2 \right\rangle$ and $T_2=\left(\frac{\sigma \tau^{3/2}}{F^{1/2}} \right)^2$.
In figure \ref{T1egalT2}, we plot $T_1$ as a function of $T_2$ for different values of $\sigma$, $\tau$, $I$ and $F$. We show that these two terms are identical and thus stationarity hypothesis is true when $\tilde{t}_{max} \gg 1$).

\bibliography{bibi}

\begin{thebibliography}{35}%
\makeatletter
\providecommand \@ifxundefined [1]{%
 \@ifx{#1\undefined}
}%
\providecommand \@ifnum [1]{%
 \ifnum #1\expandafter \@firstoftwo
 \else \expandafter \@secondoftwo
 \fi
}%
\providecommand \@ifx [1]{%
 \ifx #1\expandafter \@firstoftwo
 \else \expandafter \@secondoftwo
 \fi
}%
\providecommand \natexlab [1]{#1}%
\providecommand \enquote  [1]{``#1''}%
\providecommand \bibnamefont  [1]{#1}%
\providecommand \bibfnamefont [1]{#1}%
\providecommand \citenamefont [1]{#1}%
\providecommand \href@noop [0]{\@secondoftwo}%
\providecommand \href [0]{\begingroup \@sanitize@url \@href}%
\providecommand \@href[1]{\@@startlink{#1}\@@href}%
\providecommand \@@href[1]{\endgroup#1\@@endlink}%
\providecommand \@sanitize@url [0]{\catcode `\\12\catcode `\$12\catcode
  `\&12\catcode `\#12\catcode `\^12\catcode `\_12\catcode `\%12\relax}%
\providecommand \@@startlink[1]{}%
\providecommand \@@endlink[0]{}%
\providecommand \url  [0]{\begingroup\@sanitize@url \@url }%
\providecommand \@url [1]{\endgroup\@href {#1}{\urlprefix }}%
\providecommand \urlprefix  [0]{URL }%
\providecommand \Eprint [0]{\href }%
\providecommand \doibase [0]{http://dx.doi.org/}%
\providecommand \selectlanguage [0]{\@gobble}%
\providecommand \bibinfo  [0]{\@secondoftwo}%
\providecommand \bibfield  [0]{\@secondoftwo}%
\providecommand \translation [1]{[#1]}%
\providecommand \BibitemOpen [0]{}%
\providecommand \bibitemStop [0]{}%
\providecommand \bibitemNoStop [0]{.\EOS\space}%
\providecommand \EOS [0]{\spacefactor3000\relax}%
\providecommand \BibitemShut  [1]{\csname bibitem#1\endcsname}%
\let\auto@bib@innerbib\@empty
\bibitem [{\citenamefont {Vicsek}\ and\ \citenamefont
  {Zafeiris}(2012)}]{Vicsek2012}%
  \BibitemOpen
  \bibfield  {author} {\bibinfo {author} {\bibfnamefont {T.}~\bibnamefont
  {Vicsek}}\ and\ \bibinfo {author} {\bibfnamefont {A.}~\bibnamefont
  {Zafeiris}},\ }\href@noop {} {\bibfield  {journal} {\bibinfo  {journal}
  {Physics Reports}\ }\textbf {\bibinfo {volume} {517}},\ \bibinfo {pages} {71}
  (\bibinfo {year} {2012})}\BibitemShut {NoStop}%
\bibitem [{\citenamefont {Jacob}\ \emph {et~al.}(2000)\citenamefont {Jacob},
  \citenamefont {Cohen},\ and\ \citenamefont {Levine}}]{Levine2000}%
  \BibitemOpen
  \bibfield  {author} {\bibinfo {author} {\bibfnamefont {E.~B.}\ \bibnamefont
  {Jacob}}, \bibinfo {author} {\bibfnamefont {I.}~\bibnamefont {Cohen}}, \ and\
  \bibinfo {author} {\bibfnamefont {H.}~\bibnamefont {Levine}},\ }\href@noop {}
  {\bibfield  {journal} {\bibinfo  {journal} {Advances in Physics}\ }\textbf
  {\bibinfo {volume} {49}},\ \bibinfo {pages} {395} (\bibinfo {year}
  {2000})}\BibitemShut {NoStop}%
\bibitem [{\citenamefont {Vincenti}\ \emph {et~al.}(2019)\citenamefont
  {Vincenti}, \citenamefont {Ramos}, \citenamefont {Cordero}, \citenamefont
  {Douarche}, \citenamefont {Soto},\ and\ \citenamefont
  {Clement}}]{Clement2019}%
  \BibitemOpen
  \bibfield  {author} {\bibinfo {author} {\bibfnamefont {B.}~\bibnamefont
  {Vincenti}}, \bibinfo {author} {\bibfnamefont {G.}~\bibnamefont {Ramos}},
  \bibinfo {author} {\bibfnamefont {M.-L.}\ \bibnamefont {Cordero}}, \bibinfo
  {author} {\bibfnamefont {C.}~\bibnamefont {Douarche}}, \bibinfo {author}
  {\bibfnamefont {R.}~\bibnamefont {Soto}}, \ and\ \bibinfo {author}
  {\bibfnamefont {E.}~\bibnamefont {Clement}},\ }\href@noop {} {\bibfield
  {journal} {\bibinfo  {journal} {Nat. Comm.}\ }\textbf {\bibinfo {volume}
  {10}},\ \bibinfo {pages} {5082} (\bibinfo {year} {2019})}\BibitemShut
  {NoStop}%
\bibitem [{\citenamefont {Paxton}\ \emph {et~al.}(2005)\citenamefont {Paxton},
  \citenamefont {Sen},\ and\ \citenamefont {Mallouk}}]{Paxton2005}%
  \BibitemOpen
  \bibfield  {author} {\bibinfo {author} {\bibfnamefont {W.}~\bibnamefont
  {Paxton}}, \bibinfo {author} {\bibfnamefont {A.}~\bibnamefont {Sen}}, \ and\
  \bibinfo {author} {\bibfnamefont {T.}~\bibnamefont {Mallouk}},\ }\href@noop
  {} {\bibfield  {journal} {\bibinfo  {journal} {Eur. J. Chem.}\ }\textbf
  {\bibinfo {volume} {11}},\ \bibinfo {pages} {6462} (\bibinfo {year}
  {2005})}\BibitemShut {NoStop}%
\bibitem [{\citenamefont {Lobaskin}\ and\ \citenamefont
  {Romenskyy}(2013)}]{lobaskin2013}%
  \BibitemOpen
  \bibfield  {author} {\bibinfo {author} {\bibfnamefont {V.}~\bibnamefont
  {Lobaskin}}\ and\ \bibinfo {author} {\bibfnamefont {M.}~\bibnamefont
  {Romenskyy}},\ }\href@noop {} {\bibfield  {journal} {\bibinfo  {journal}
  {Phys. Rev. E}\ }\textbf {\bibinfo {volume} {87}},\ \bibinfo {pages} {052135}
  (\bibinfo {year} {2013})}\BibitemShut {NoStop}%
\bibitem [{\citenamefont {Gueron}\ and\ \citenamefont
  {A.Levin}(1993)}]{Levin2019}%
  \BibitemOpen
  \bibfield  {author} {\bibinfo {author} {\bibfnamefont {S.}~\bibnamefont
  {Gueron}}\ and\ \bibinfo {author} {\bibfnamefont {S.}~\bibnamefont
  {A.Levin}},\ }\href@noop {} {\bibfield  {journal} {\bibinfo  {journal} {J. Of
  Theor. Biol.}\ }\textbf {\bibinfo {volume} {165}},\ \bibinfo {pages} {541}
  (\bibinfo {year} {1993})}\BibitemShut {NoStop}%
\bibitem [{\citenamefont {Radakov}(1973)}]{Radakov1973}%
  \BibitemOpen
  \bibfield  {author} {\bibinfo {author} {\bibfnamefont {D.~V.}\ \bibnamefont
  {Radakov}},\ }\href@noop {} {\emph {\bibinfo {title} {Schooling in the
  Ecology of Fish}}}\ (\bibinfo  {publisher} {John Wiley New York},\ \bibinfo
  {year} {1973})\BibitemShut {NoStop}%
\bibitem [{\citenamefont {Pavlov}\ and\ \citenamefont
  {Kasumyan}(2000)}]{Pavlov2000}%
  \BibitemOpen
  \bibfield  {author} {\bibinfo {author} {\bibfnamefont {D.}~\bibnamefont
  {Pavlov}}\ and\ \bibinfo {author} {\bibfnamefont {A.~O.}\ \bibnamefont
  {Kasumyan}},\ }\href@noop {} {\bibfield  {journal} {\bibinfo  {journal}
  {Journal of Ichthyology}\ }\textbf {\bibinfo {volume} {40}},\ \bibinfo
  {pages} {S163} (\bibinfo {year} {2000})}\BibitemShut {NoStop}%
\bibitem [{\citenamefont {Liao}(2007)}]{Liao2007}%
  \BibitemOpen
  \bibfield  {author} {\bibinfo {author} {\bibfnamefont {J.}~\bibnamefont
  {Liao}},\ }\href@noop {} {\bibfield  {journal} {\bibinfo  {journal} {Philos
  Trans R Soc Lond B Biol Sci.}\ }\textbf {\bibinfo {volume} {362}},\ \bibinfo
  {pages} {1973} (\bibinfo {year} {2007})}\BibitemShut {NoStop}%
\bibitem [{\citenamefont {Larsson}(2012)}]{Larsson2012}%
  \BibitemOpen
  \bibfield  {author} {\bibinfo {author} {\bibfnamefont {M.}~\bibnamefont
  {Larsson}},\ }\href@noop {} {\bibfield  {journal} {\bibinfo  {journal}
  {Current Zoology}\ }\textbf {\bibinfo {volume} {58}},\ \bibinfo {pages} {116}
  (\bibinfo {year} {2012})}\BibitemShut {NoStop}%
\bibitem [{\citenamefont {Parr}(1927)}]{Parr1927}%
  \BibitemOpen
  \bibfield  {author} {\bibinfo {author} {\bibfnamefont {A.~E.}\ \bibnamefont
  {Parr}},\ }\href@noop {} {\bibfield  {journal} {\bibinfo  {journal} {Occas.
  Pap. Bigham Ocean- ogr. Coil.}\ }\textbf {\bibinfo {volume} {1}},\ \bibinfo
  {pages} {1} (\bibinfo {year} {1927})}\BibitemShut {NoStop}%
\bibitem [{\citenamefont {Breder}(1951)}]{Breder1951}%
  \BibitemOpen
  \bibfield  {author} {\bibinfo {author} {\bibfnamefont {C.~M.}\ \bibnamefont
  {Breder}},\ }\href@noop {} {\bibfield  {journal} {\bibinfo  {journal}
  {Bulletin of the American Museum of Natural History}\ }\textbf {\bibinfo
  {volume} {98}},\ \bibinfo {pages} {1} (\bibinfo {year} {1951})}\BibitemShut
  {NoStop}%
\bibitem [{\citenamefont {Svendsen}\ \emph {et~al.}(2003)\citenamefont
  {Svendsen}, \citenamefont {Bildsoe},\ and\ \citenamefont
  {Steffensen}}]{Svendsen2003}%
  \BibitemOpen
  \bibfield  {author} {\bibinfo {author} {\bibfnamefont {J.~C.}\ \bibnamefont
  {Svendsen}}, \bibinfo {author} {\bibfnamefont {M.}~\bibnamefont {Bildsoe}}, \
  and\ \bibinfo {author} {\bibfnamefont {J.}~\bibnamefont {Steffensen}},\
  }\href@noop {} {\bibfield  {journal} {\bibinfo  {journal} {J. Fish Biol.}\
  }\textbf {\bibinfo {volume} {62}},\ \bibinfo {pages} {834} (\bibinfo {year}
  {2003})}\BibitemShut {NoStop}%
\bibitem [{\citenamefont {Ross}\ and\ \citenamefont
  {Backman}(1992)}]{Ross1992}%
  \BibitemOpen
  \bibfield  {author} {\bibinfo {author} {\bibfnamefont {R.~M.}\ \bibnamefont
  {Ross}}\ and\ \bibinfo {author} {\bibfnamefont {T.~W.~H.}\ \bibnamefont
  {Backman}},\ }\href@noop {} {\bibfield  {journal} {\bibinfo  {journal}
  {American shad. Trans. Am. Fish. Soc.}\ }\textbf {\bibinfo {volume} {121}},\
  \bibinfo {pages} {385} (\bibinfo {year} {1992})}\BibitemShut {NoStop}%
\bibitem [{\citenamefont {Ashrafa}\ \emph {et~al.}(2017)\citenamefont
  {Ashrafa}, \citenamefont {Bradshawa}, \citenamefont {Haa}, \citenamefont
  {Halloyb}, \citenamefont {Godoy-Diana},\ and\ \citenamefont
  {Thiria}}]{Ashrafa2017}%
  \BibitemOpen
  \bibfield  {author} {\bibinfo {author} {\bibfnamefont {I.}~\bibnamefont
  {Ashrafa}}, \bibinfo {author} {\bibfnamefont {H.}~\bibnamefont {Bradshawa}},
  \bibinfo {author} {\bibfnamefont {T.-T.}\ \bibnamefont {Haa}}, \bibinfo
  {author} {\bibfnamefont {J.}~\bibnamefont {Halloyb}}, \bibinfo {author}
  {\bibfnamefont {R.}~\bibnamefont {Godoy-Diana}}, \ and\ \bibinfo {author}
  {\bibfnamefont {B.}~\bibnamefont {Thiria}},\ }\href@noop {} {\bibfield
  {journal} {\bibinfo  {journal} {PNAS}\ }\textbf {\bibinfo {volume} {114}},\
  \bibinfo {pages} {9599} (\bibinfo {year} {2017})}\BibitemShut {NoStop}%
\bibitem [{\citenamefont {Parker}(1973)}]{Parker1973}%
  \BibitemOpen
  \bibfield  {author} {\bibinfo {author} {\bibfnamefont {F.~R.~J.}\
  \bibnamefont {Parker}},\ }\href@noop {} {\bibfield  {journal} {\bibinfo
  {journal} {Trans. Am. Fish. Soc.}\ }\textbf {\bibinfo {volume} {102}},\
  \bibinfo {pages} {125} (\bibinfo {year} {1973})}\BibitemShut {NoStop}%
\bibitem [{\citenamefont {Partridge}\ and\ \citenamefont
  {Pitcher}(1980)}]{Partridge1980}%
  \BibitemOpen
  \bibfield  {author} {\bibinfo {author} {\bibfnamefont {B.~L.}\ \bibnamefont
  {Partridge}}\ and\ \bibinfo {author} {\bibfnamefont {T.}~\bibnamefont
  {Pitcher}},\ }\href@noop {} {\bibfield  {journal} {\bibinfo  {journal} {J.
  Comp. Physiol. A.}\ }\textbf {\bibinfo {volume} {135}},\ \bibinfo {pages}
  {315} (\bibinfo {year} {1980})}\BibitemShut {NoStop}%
\bibitem [{\citenamefont {Krause}(1993)}]{Krause1993}%
  \BibitemOpen
  \bibfield  {author} {\bibinfo {author} {\bibfnamefont {J.}~\bibnamefont
  {Krause}},\ }\href@noop {} {\bibfield  {journal} {\bibinfo  {journal}
  {Oecologia}\ }\textbf {\bibinfo {volume} {93}},\ \bibinfo {pages} {356}
  (\bibinfo {year} {1993})}\BibitemShut {NoStop}%
\bibitem [{\citenamefont {Reichhardt}\ and\ \citenamefont
  {Reichhardt}(2018)}]{Reichhardt2018}%
  \BibitemOpen
  \bibfield  {author} {\bibinfo {author} {\bibfnamefont {C.}~\bibnamefont
  {Reichhardt}}\ and\ \bibinfo {author} {\bibfnamefont {C.}~\bibnamefont
  {Reichhardt}},\ }\href@noop {} {\bibfield  {journal} {\bibinfo  {journal}
  {Soft Matter}\ }\textbf {\bibinfo {volume} {14}},\ \bibinfo {pages} {490}
  (\bibinfo {year} {2018})}\BibitemShut {NoStop}%
\bibitem [{\citenamefont {Vicsek}\ \emph {et~al.}(1995)\citenamefont {Vicsek},
  \citenamefont {Czirok}, \citenamefont {Ben-Jacob}, \citenamefont {Cohen},\
  and\ \citenamefont {Shochet}}]{Vicsek1995}%
  \BibitemOpen
  \bibfield  {author} {\bibinfo {author} {\bibfnamefont {T.}~\bibnamefont
  {Vicsek}}, \bibinfo {author} {\bibfnamefont {A.}~\bibnamefont {Czirok}},
  \bibinfo {author} {\bibfnamefont {E.}~\bibnamefont {Ben-Jacob}}, \bibinfo
  {author} {\bibfnamefont {I.}~\bibnamefont {Cohen}}, \ and\ \bibinfo {author}
  {\bibfnamefont {O.}~\bibnamefont {Shochet}},\ }\href@noop {} {\bibfield
  {journal} {\bibinfo  {journal} {Phys. rev. Lett.}\ }\textbf {\bibinfo
  {volume} {75}},\ \bibinfo {pages} {1226} (\bibinfo {year}
  {1995})}\BibitemShut {NoStop}%
\bibitem [{\citenamefont {Gautrais}\ \emph {et~al.}(2009)\citenamefont
  {Gautrais}, \citenamefont {Jost}, \citenamefont {Soria}, \citenamefont
  {Campo}, \citenamefont {Motsch}, \citenamefont {Fournier}, \citenamefont
  {Blanco},\ and\ \citenamefont {Theraulaz}}]{Gautrais2009}%
  \BibitemOpen
  \bibfield  {author} {\bibinfo {author} {\bibfnamefont {J.}~\bibnamefont
  {Gautrais}}, \bibinfo {author} {\bibfnamefont {C.}~\bibnamefont {Jost}},
  \bibinfo {author} {\bibfnamefont {M.}~\bibnamefont {Soria}}, \bibinfo
  {author} {\bibfnamefont {A.}~\bibnamefont {Campo}}, \bibinfo {author}
  {\bibfnamefont {S.}~\bibnamefont {Motsch}}, \bibinfo {author} {\bibfnamefont
  {R.}~\bibnamefont {Fournier}}, \bibinfo {author} {\bibfnamefont
  {S.}~\bibnamefont {Blanco}}, \ and\ \bibinfo {author} {\bibfnamefont
  {G.}~\bibnamefont {Theraulaz}},\ }\href@noop {} {\bibfield  {journal}
  {\bibinfo  {journal} {J. Math. Biol.}\ }\textbf {\bibinfo {volume} {58}},\
  \bibinfo {pages} {429} (\bibinfo {year} {2009})}\BibitemShut {NoStop}%
\bibitem [{\citenamefont {Gautrais}\ \emph {et~al.}(2012)\citenamefont
  {Gautrais}, \citenamefont {Ginelli}, \citenamefont {Fournier}, \citenamefont
  {Blanco}, \citenamefont {Soria}, \citenamefont {Chat\'e},\ and\ \citenamefont
  {Theraulaz}}]{Gautrais2012}%
  \BibitemOpen
  \bibfield  {author} {\bibinfo {author} {\bibfnamefont {J.}~\bibnamefont
  {Gautrais}}, \bibinfo {author} {\bibfnamefont {F.}~\bibnamefont {Ginelli}},
  \bibinfo {author} {\bibfnamefont {R.}~\bibnamefont {Fournier}}, \bibinfo
  {author} {\bibfnamefont {S.}~\bibnamefont {Blanco}}, \bibinfo {author}
  {\bibfnamefont {M.}~\bibnamefont {Soria}}, \bibinfo {author} {\bibfnamefont
  {H.}~\bibnamefont {Chat\'e}}, \ and\ \bibinfo {author} {\bibfnamefont
  {G.}~\bibnamefont {Theraulaz}},\ }\href@noop {} {\bibfield  {journal}
  {\bibinfo  {journal} {PLoS Comput. Biol.}\ }\textbf {\bibinfo {volume} {8}},\
  \bibinfo {pages} {e1002678} (\bibinfo {year} {2012})}\BibitemShut {NoStop}%
\bibitem [{\citenamefont {Calovi}\ \emph {et~al.}(2014)\citenamefont {Calovi},
  \citenamefont {Lopez}, \citenamefont {Ngo}, \citenamefont {Sire},
  \citenamefont {Chat\'e},\ and\ \citenamefont {Theraulaz}}]{Calovi2014}%
  \BibitemOpen
  \bibfield  {author} {\bibinfo {author} {\bibfnamefont {D.~S.}\ \bibnamefont
  {Calovi}}, \bibinfo {author} {\bibfnamefont {U.}~\bibnamefont {Lopez}},
  \bibinfo {author} {\bibfnamefont {S.}~\bibnamefont {Ngo}}, \bibinfo {author}
  {\bibfnamefont {C.}~\bibnamefont {Sire}}, \bibinfo {author} {\bibfnamefont
  {H.}~\bibnamefont {Chat\'e}}, \ and\ \bibinfo {author} {\bibfnamefont
  {G.}~\bibnamefont {Theraulaz}},\ }\href@noop {} {\bibfield  {journal}
  {\bibinfo  {journal} {New Journal of Physics}\ }\textbf {\bibinfo {volume}
  {16}},\ \bibinfo {pages} {015026} (\bibinfo {year} {2014})}\BibitemShut
  {NoStop}%
\bibitem [{\citenamefont {S.Calovi}\ \emph {et~al.}(2018)\citenamefont
  {S.Calovi}, \citenamefont {Litchinko}, \citenamefont {Lecheval},
  \citenamefont {Lopez}, \citenamefont {Escudero}, \citenamefont {Chat\'e},
  \citenamefont {Sire},\ and\ \citenamefont {Theraulaz}}]{Calovi2018}%
  \BibitemOpen
  \bibfield  {author} {\bibinfo {author} {\bibfnamefont {D.}~\bibnamefont
  {S.Calovi}}, \bibinfo {author} {\bibfnamefont {A.}~\bibnamefont {Litchinko}},
  \bibinfo {author} {\bibfnamefont {V.}~\bibnamefont {Lecheval}}, \bibinfo
  {author} {\bibfnamefont {U.}~\bibnamefont {Lopez}}, \bibinfo {author}
  {\bibfnamefont {A.~P.}\ \bibnamefont {Escudero}}, \bibinfo {author}
  {\bibfnamefont {H.}~\bibnamefont {Chat\'e}}, \bibinfo {author} {\bibfnamefont
  {C.}~\bibnamefont {Sire}}, \ and\ \bibinfo {author} {\bibfnamefont
  {G.}~\bibnamefont {Theraulaz}},\ }\href@noop {} {\bibfield  {journal}
  {\bibinfo  {journal} {PLOS Computational Biology}\ }\textbf {\bibinfo
  {volume} {14}},\ \bibinfo {pages} {e1005933} (\bibinfo {year}
  {2018})}\BibitemShut {NoStop}%
\bibitem [{\citenamefont {Hamilton}(1971)}]{Hamilton1971}%
  \BibitemOpen
  \bibfield  {author} {\bibinfo {author} {\bibfnamefont {W.~D.}\ \bibnamefont
  {Hamilton}},\ }\href@noop {} {\bibfield  {journal} {\bibinfo  {journal} {J.
  theor. Biol.}\ }\textbf {\bibinfo {volume} {31}},\ \bibinfo {pages} {295}
  (\bibinfo {year} {1971})}\BibitemShut {NoStop}%
\bibitem [{\citenamefont {H.A.Keenleyside}(1979)}]{Keenleyside1979}%
  \BibitemOpen
  \bibfield  {author} {\bibinfo {author} {\bibfnamefont {M.}~\bibnamefont
  {H.A.Keenleyside}},\ }\href@noop {} {\emph {\bibinfo {title} {Diversity and
  Adaptation in Fish Behaviour}}},\ Vol.~\bibinfo {volume} {11}\ (\bibinfo
  {publisher} {Springer Verlan, New York 1979},\ \bibinfo {year} {1979})\ p.\
  \bibinfo {pages} {170}\BibitemShut {NoStop}%
\bibitem [{\citenamefont {Potts}(1970)}]{Potts1970}%
  \BibitemOpen
  \bibfield  {author} {\bibinfo {author} {\bibfnamefont {G.~W.}\ \bibnamefont
  {Potts}},\ }\href@noop {} {\bibfield  {journal} {\bibinfo  {journal} {J.
  Zool.}\ }\textbf {\bibinfo {volume} {161}},\ \bibinfo {pages} {223} (\bibinfo
  {year} {1970})}\BibitemShut {NoStop}%
\bibitem [{\citenamefont {Montgomery}\ \emph {et~al.}(1997)\citenamefont
  {Montgomery}, \citenamefont {Baker},\ and\ \citenamefont
  {Carton}}]{Montgomery1997}%
  \BibitemOpen
  \bibfield  {author} {\bibinfo {author} {\bibfnamefont {J.~C.}\ \bibnamefont
  {Montgomery}}, \bibinfo {author} {\bibfnamefont {C.~F.}\ \bibnamefont
  {Baker}}, \ and\ \bibinfo {author} {\bibfnamefont {A.~G.}\ \bibnamefont
  {Carton}},\ }\href@noop {} {\bibfield  {journal} {\bibinfo  {journal}
  {Nature}\ }\textbf {\bibinfo {volume} {389}},\ \bibinfo {pages} {960}
  (\bibinfo {year} {1997})}\BibitemShut {NoStop}%
\bibitem [{\citenamefont {Kulpa}\ \emph {et~al.}(2015)\citenamefont {Kulpa},
  \citenamefont {Bak-Coleman},\ and\ \citenamefont {Coombs}}]{Kulpa2015}%
  \BibitemOpen
  \bibfield  {author} {\bibinfo {author} {\bibfnamefont {M.}~\bibnamefont
  {Kulpa}}, \bibinfo {author} {\bibfnamefont {J.}~\bibnamefont {Bak-Coleman}},
  \ and\ \bibinfo {author} {\bibfnamefont {S.}~\bibnamefont {Coombs}},\
  }\href@noop {} {\bibfield  {journal} {\bibinfo  {journal} {The Journal of
  Experimental Biology}\ }\textbf {\bibinfo {volume} {218}},\ \bibinfo {pages}
  {1603} (\bibinfo {year} {2015})}\BibitemShut {NoStop}%
\bibitem [{\citenamefont {Simmonds}\ and\ \citenamefont
  {MacLennan}(2008)}]{Simmonds2008}%
  \BibitemOpen
  \bibfield  {author} {\bibinfo {author} {\bibfnamefont {J.}~\bibnamefont
  {Simmonds}}\ and\ \bibinfo {author} {\bibfnamefont {D.~N.}\ \bibnamefont
  {MacLennan}},\ }\href@noop {} {\emph {\bibinfo {title} {Fisheries Acoustics:
  Theory and Practices}}}\ (\bibinfo  {publisher} {Wiley 2008},\ \bibinfo
  {year} {2008})\BibitemShut {NoStop}%
\bibitem [{\citenamefont {Helbing}\ and\ \citenamefont
  {Moln\'ar}(1995)}]{Helbing1995}%
  \BibitemOpen
  \bibfield  {author} {\bibinfo {author} {\bibfnamefont {D.}~\bibnamefont
  {Helbing}}\ and\ \bibinfo {author} {\bibfnamefont {P.}~\bibnamefont
  {Moln\'ar}},\ }\href@noop {} {\bibfield  {journal} {\bibinfo  {journal}
  {Phys. Rev. E}\ }\textbf {\bibinfo {volume} {51}},\ \bibinfo {pages} {4282}
  (\bibinfo {year} {1995})}\BibitemShut {NoStop}%
\bibitem [{\citenamefont {Helbing}\ \emph {et~al.}(2000)\citenamefont
  {Helbing}, \citenamefont {Farkas},\ and\ \citenamefont
  {Vicsek}}]{Helbing2000}%
  \BibitemOpen
  \bibfield  {author} {\bibinfo {author} {\bibfnamefont {D.}~\bibnamefont
  {Helbing}}, \bibinfo {author} {\bibfnamefont {I.}~\bibnamefont {Farkas}}, \
  and\ \bibinfo {author} {\bibfnamefont {T.}~\bibnamefont {Vicsek}},\
  }\href@noop {} {\bibfield  {journal} {\bibinfo  {journal} {Nature}\ }\textbf
  {\bibinfo {volume} {407}},\ \bibinfo {pages} {487} (\bibinfo {year}
  {2000})}\BibitemShut {NoStop}%
\bibitem [{\citenamefont {Porter}\ \emph {et~al.}(2011)\citenamefont {Porter},
  \citenamefont {Roque},\ and\ \citenamefont {Jr}}]{Porter2011}%
  \BibitemOpen
  \bibfield  {author} {\bibinfo {author} {\bibfnamefont {M.~E.}\ \bibnamefont
  {Porter}}, \bibinfo {author} {\bibfnamefont {C.~M.}\ \bibnamefont {Roque}}, \
  and\ \bibinfo {author} {\bibfnamefont {J.~H.~L.}\ \bibnamefont {Jr}},\
  }\href@noop {} {\bibfield  {journal} {\bibinfo  {journal} {Zoology}\ }\textbf
  {\bibinfo {volume} {114}},\ \bibinfo {pages} {348} (\bibinfo {year}
  {2011})}\BibitemShut {NoStop}%
\bibitem [{\citenamefont {Higham}(2001)}]{Higham2001}%
  \BibitemOpen
  \bibfield  {author} {\bibinfo {author} {\bibfnamefont {D.~J.}\ \bibnamefont
  {Higham}},\ }\href@noop {} {\bibfield  {journal} {\bibinfo  {journal} {SIAM
  Review}\ }\textbf {\bibinfo {volume} {43}},\ \bibinfo {pages} {525} (\bibinfo
  {year} {2001})}\BibitemShut {NoStop}%
\bibitem [{\citenamefont {Oteiza}\ \emph {et~al.}(2017)\citenamefont {Oteiza},
  \citenamefont {Odstril}, \citenamefont {Lauder}, \citenamefont {Portugues},\
  and\ \citenamefont {Engert}}]{Oteiza2017}%
  \BibitemOpen
  \bibfield  {author} {\bibinfo {author} {\bibfnamefont {P.}~\bibnamefont
  {Oteiza}}, \bibinfo {author} {\bibfnamefont {I.}~\bibnamefont {Odstril}},
  \bibinfo {author} {\bibfnamefont {G.}~\bibnamefont {Lauder}}, \bibinfo
  {author} {\bibfnamefont {R.}~\bibnamefont {Portugues}}, \ and\ \bibinfo
  {author} {\bibfnamefont {F.}~\bibnamefont {Engert}},\ }\href@noop {}
  {\bibfield  {journal} {\bibinfo  {journal} {Nature}\ }\textbf {\bibinfo
  {volume} {547}},\ \bibinfo {pages} {445} (\bibinfo {year}
  {2017})}\BibitemShut {NoStop}%
\end{thebibliography}%

\end{document}